\documentclass[aps,prl,twocolumn,floatfix,superscriptaddress,showpacs,amsfonts,amssymb,amsmath,preprintnumbers]{revtex4}
\usepackage{bm}
\usepackage{psfig}
\usepackage{dcolumn}

\newcommand{\Eqsdash}[2]{Equations~(\ref{#1}--\ref{#2})}
\newcommand{\eqsdash}[2]{Eqs.~(\ref{#1}--\ref{#2})}

\newcommand{\exref}[1]{(\ref{#1})}

\newcommand{\Figref}[1]{Figure~\ref{#1}}
\newcommand{\figref}[1]{Fig.~\ref{#1}}

\newcommand{\bea}{\begin{eqnarray}}
\newcommand{\eea}{\end{eqnarray}}
\newcommand{\bal}{\begin{aligned}}
\newcommand{\eal}{\end{aligned}}
\newcommand{\bga}{\begin{gathered}}
\newcommand{\ega}{\end{gathered}}

\newcommand{\bl}{\bigl}
\newcommand{\br}{\bigr}
\newcommand{\la}{\langle}
\newcommand{\ra}{\rangle}

\newcommand{\const}{\text{const}}
\renewcommand{\phi}{\varphi}

\newcommand{\dd}{\partial}

\newcommand{\vu}{{\bf u}}
\newcommand{\vf}{{\bf f}}
\newcommand{\vB}{{\bf B}}

\newcommand{\vx}{{\bf x}}
\newcommand{\vy}{{\bf y}}

\newcommand{\kpar}{k_\parallel}
\newcommand{\krms}{k_\text{rms}}

\newcommand{\kf}{k_0}

\newcommand{\kl}{k_\lambda}
\newcommand{\kres}{k_\eta}

\newcommand{\lf}{\ell_0}
\newcommand{\ld}{\ell_\nu}
\newcommand{\lres}{\ell_\eta}

\newcommand{\tcorr}{\tau_\text{c}}

\newcommand{\usq}{\la u^2\ra}
\newcommand{\Bmean}{\la \vB\ra}
\newcommand{\Bsq}{\la B^2\ra}
\newcommand{\Bfr}{\la B^4\ra}

\renewcommand{\Pr}{\text{Pr}_\text{m}} 
\newcommand{\Prc}{\text{Pr}_\text{m,c}} 
\renewcommand{\Re}{\text{Re}} 

\newcommand{\Rm}{\text{Rm}} 
\newcommand{\Rmc}{\text{Rm}_\text{c}}

\begin{document}

\preprint{E-print {\tt astro-ph/0308336}}

\title{Critical Magnetic Prandtl Number for Small-Scale Dynamo}

\author{Alexander A.\ Schekochihin}
\email{as629@damtp.cam.ac.uk}
\altaffiliation[Present-time address: ]{DAMTP/CMS, 
University of Cambridge, Wilberforce Rd., Cambridge CB3 0WA, UK.}
\affiliation{Plasma Physics Group, Imperial College, 
Blackett Laboratory, Prince Consort Road, London~SW7~2BW, UK}
\author{Steven C.\ Cowley}
\affiliation{Plasma Physics Group, Imperial College, 
Blackett Laboratory, Prince Consort Road, London~SW7~2BW, UK}
\affiliation{Department of Physics and Astronomy, 
UCLA, Los Angeles, CA 90095-1547}
\author{Jason L.\ Maron}
\affiliation{Department of Physics and Astronomy, University of Rochester, 
Rochester, NY 14627} 
\affiliation{Center for Magnetic Reconnection Studies, 
Department of Physics and Astronomy, University of Iowa, 
Iowa City, IA 52242}
\author{James C.\ McWilliams}
\affiliation{Department of Atmospheric Sciences, 
UCLA, Los Angeles, CA 90095-1565}
\date{\today}

\begin{abstract}
We report a series of numerical simulations 
showing that the critical magnetic Reynolds number 
$\Rmc$ for the nonhelical small-scale dynamo 
depends on the Reynolds number $\Re$. 
Namely, the dynamo is 
shut down if the magnetic Prandtl number $\Pr=\Rm/\Re$ 
is less than some critical value $\Prc\lesssim1$ 
even for $\Rm$ for which dynamo exists at $\Pr\ge1$. 
We argue that, in the limit of $\Re\to\infty$, 
a finite $\Prc$ may exist. 
The second possibility is that $\Prc\to0$ as $\Re\to\infty$, 
while $\Rmc$ tends to a very large constant value 
inaccessible at current resolutions. 
If there is a finite $\Prc$, the dynamo is sustainable only 
if magnetic fields 
can exist at scales smaller than the flow scale, 
i.e., it is always effectively a large-$\Pr$ dynamo. 
If there is a finite $\Rmc$, our results provide 
a lower bound: $\Rmc\gtrsim220$ for $\Pr\le1/8$. 
This is larger than $\Rm$ in many planets 
and in all liquid-metal experiments. 
\end{abstract}


\maketitle

The simplest description of a conducting fluid is in terms of 
equations of magnetohydrodynamics (MHD): 
\bea
\label{NSEq}
\dd_t\vu + \vu\cdot\nabla\vu &=& \nu\Delta\vu - \nabla p + \vB\cdot\nabla\vB + \vf,\\
\label{ind_eq}
\dd_t\vB + \vu\cdot\nabla\vB &=& \vB\cdot\nabla\vu + \eta\Delta\vB,
\eea
where 
$\vu$~is velocity, 
$\vB$~is magnetic field, 
$\vf$~is the external-force density, 
$\nu$~is viscosity and $\eta$~is magnetic diffusivity. 
The pressure gradient $\nabla p$ 
is determined by the incompressibility condition 
$\nabla\cdot\vu = 0$. We have rescaled $p$ and $\vB$ 
by~$\rho$ and~$(4\pi\rho)^{1/2}$, respectively 
($\rho$ is density). 

A fundamental property of \eqsdash{NSEq}{ind_eq} is the ability of the 
velocity and magnetic fields to exchange energy. 
In three-dimensional turbulent flows (and in many other 
chaotic flows), this can take the form of net amplification 
of magnetic field with time, a process referred to as MHD dynamo. 
There are two kinds of dynamo. The first is the mean-field 
dynamo (growth of $\la\vB\ra$), which usually 
requires a net flow helicity \cite{Moffatt_book}. 
It is a large-scale effect that must be considered 
in conjunction with such system-specific properties as geometry, 
rotation, mean shear etc. 
The second kind is the small-scale dynamo: 
amplification of the magnetic energy~$\Bsq$ 
due to random stretching of the field by the turbulent 
flow, requiring no net helicity 
\cite{Batchelor_dynamo,Zeldovich_etal_linear,STF_book}. 
The stretching is opposed by the resistive diffusion, 
so the dynamo is only possible when the magnetic Reynolds 
number $\Rm=\usq^{1/2}\lf/\eta$ exceeds a certain critical value 
($\lf$ is the system scale). 

In this Letter, we study the existence of small-scale dynamo 
in homogeneous incompressible turbulence  
with magnetic Prandtl number $\Pr=\nu/\eta<1$
(i.e., $\Re=\usq^{1/2}\lf/\nu>\Rm$). 
This is an important issue because $\Pr$ is small in 
stars ($\Pr\sim10^{-2}$ at base of the Sun's convection zone), 
planets ($\Pr\sim10^{-5}$ \cite{Roberts_Glatzmaier_review}), 
and in liquid-metal laboratory dynamos 
\cite{Gailitis_etal_review,Forest_etal,Bourgoin_etal}. 

In three dimensions, most types of turbulence 
at scales much smaller than the system size are predominantly 
vortical and well described by Kolmogorov's dimensional 
theory \cite{Frisch_book}. In this theory, the fastest field 
stretching is done by the small-scale velocities. 
The essential physics of small-scale dynamo should thus be contained 
within our homogeneous, isotropic, incompressible model. 

\begin{figure*}[t]
\centerline{\psfig{file=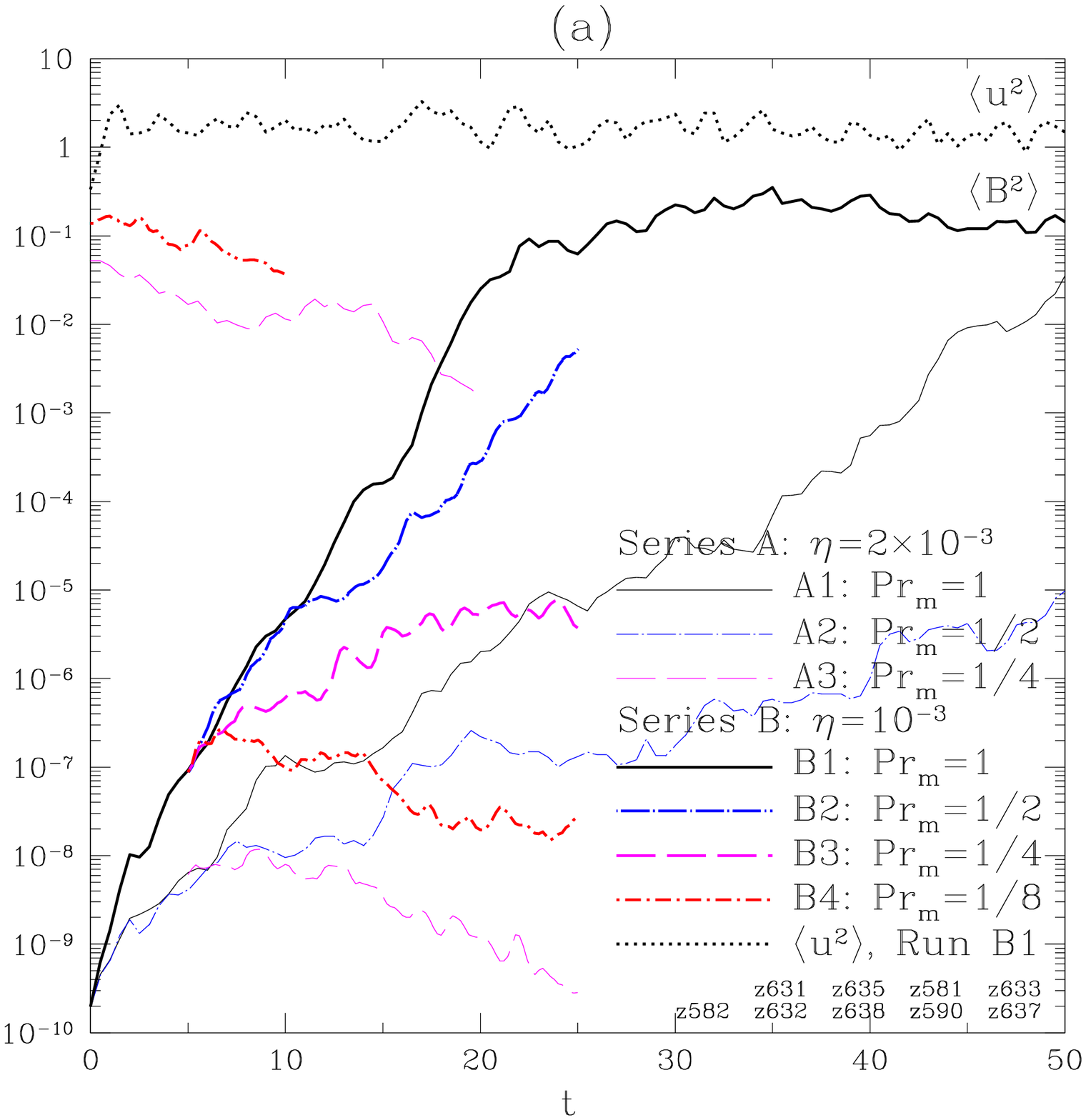,width=8.25cm}
\psfig{file=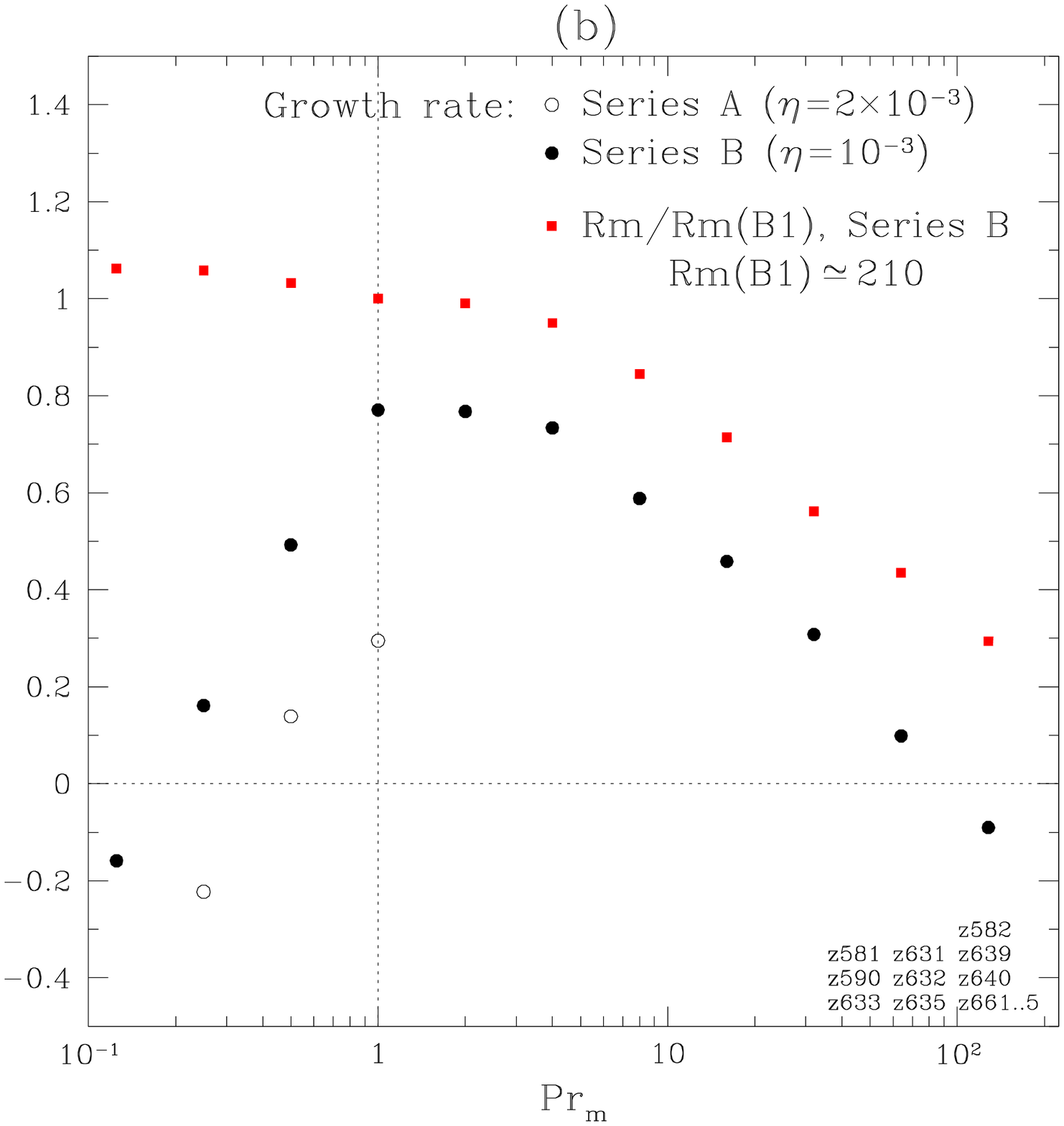,width=8.25cm}}
\caption{\label{fig_Et} (a) Evolution of magnetic energy~$\Bsq(t)$ 
in all runs. 
Also shown are the versions of Runs~A3 and~B4 
that started from the saturated state of Run~A1 (at $t=75$) and 
Run~B1 (at $t=50$), respectively. 
Runs~A1, A2, and B1 have resolution $128^3$, 
Runs~A3, B2, B3, and B4 are $256^3$. 
(b) Growth rates vs.~$\Pr$ (at fixed~$\eta$) 
for the same runs plus for 7 runs 
($128^3$) extending Series~B to $\Pr>1$. 
The dynamo is again shut down at $\Pr\sim100$ because velocity 
is strongly damped by large viscosity and $\Rm$ drops below 
critical. We show $\Rm$ (divided 
by its value for Run B1, $\Rm\simeq210$) 
in the same plot.} 
\end{figure*}

The small-scale dynamo is most transparent 
in the limit of $\Pr\gg1$. 
Straightforward estimates show that, while velocity 
is dissipated at the viscous scale~$\ld\sim\Re^{-3/4}\lf$, 
magnetic field can occupy smaller scales 
down to the resistive $\lres\sim\Pr^{-1/2}\ld$. 
The dynamo is driven by the fastest eddies: the 
viscous-scale ones, which are spatially smooth. 
The growing fields are organized in folds, with direction reversals 
at the resistive scale~$\lres$ and field lines remaining 
approximately straight up to the flow scale~$\ld$ 
\cite{Ott_review,Chertkov_etal_dynamo,SCMM_folding,SCTMM_stokes}. 
Why this is a winning configuration is best seen 
on the example of a linear velocity field 
(locally uniform rate of strain) 
\cite{Zeldovich_etal_linear,Chertkov_etal_dynamo}: 
the field aligns with the stretching direction of the 
flow but reverses along the ``null'' direction, 
so that compression cannot lead to resistive 
annihilation of antiparallel fields canceling the 
effect of stretching. For this mechanism to apply, 
it is essential that (i) the flow be spatially 
smooth, so fluid trajectories separate exponentially in time 
leading to exponential stretching, (ii) a scale separation 
between the field scale (reversals) and the flow be 
achievable. 
The large-$\Pr$ turbulent dynamo satisfies both conditions, 
as do all deterministic chaotic dynamos \cite{STF_book}. 
Thus, the small-scale dynamo, as it is usually understood, 
is essentially the large-$\Pr$ dynamo. 
The often simulated case of $\Pr=1$ belongs to 
the same class: the magnetic energy is amplified at scales 
somewhat smaller than the viscous scale and the field structure is 
similar to the case of~$\Pr\gg1$ \cite{SCTMM_stokes}. 
When $\Pr<1$, the field scale is resistively 
limited to be comparable to, or larger than, the viscous cutoff. 
The field interacts with inertial-range motions, which 
are spatially rough and cannot be thought of as having a locally 
uniform rate of strain. Is there still a small-scale dynamo? 

\begin{figure*}[t]
\centerline{\psfig{file=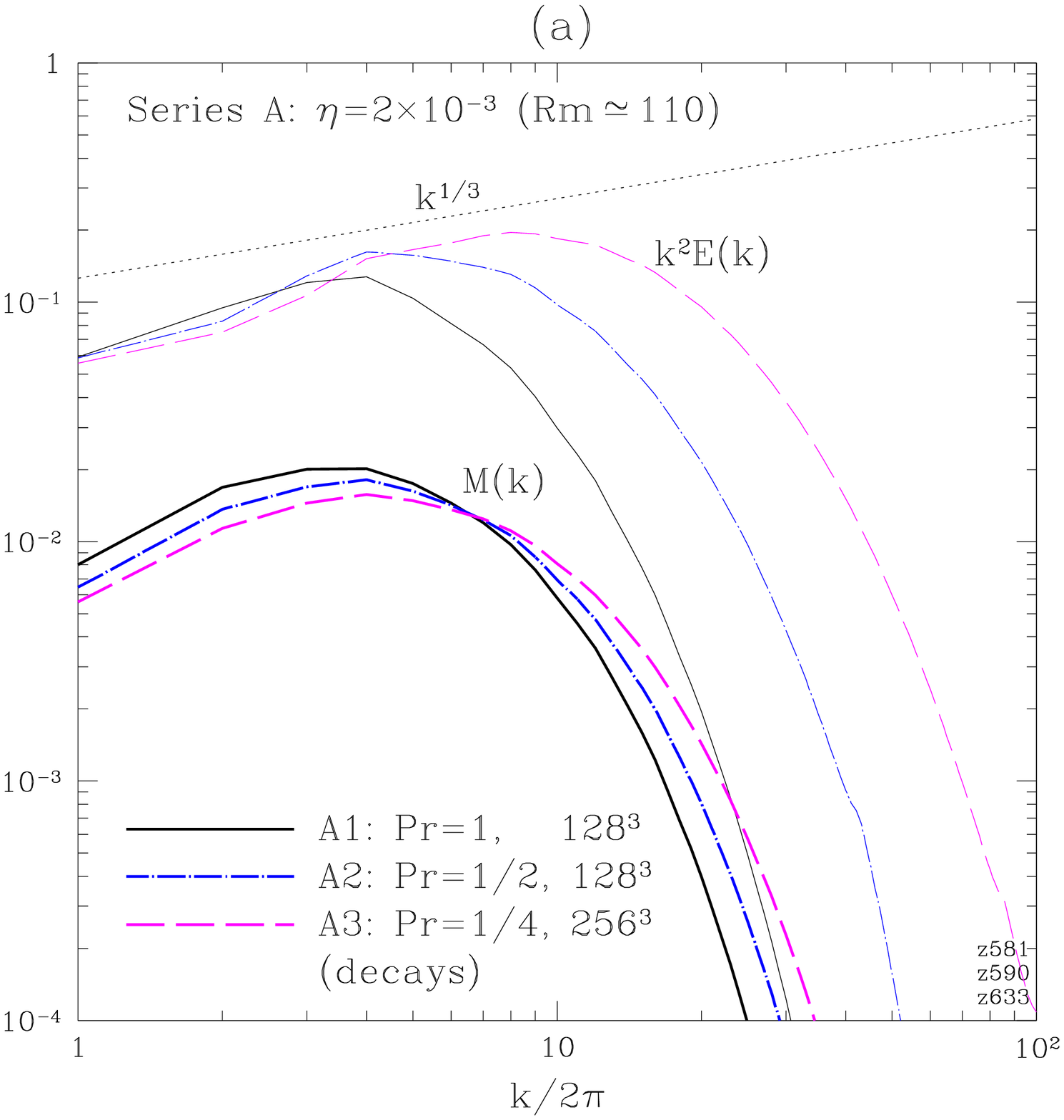,width=8.25cm}
\psfig{file=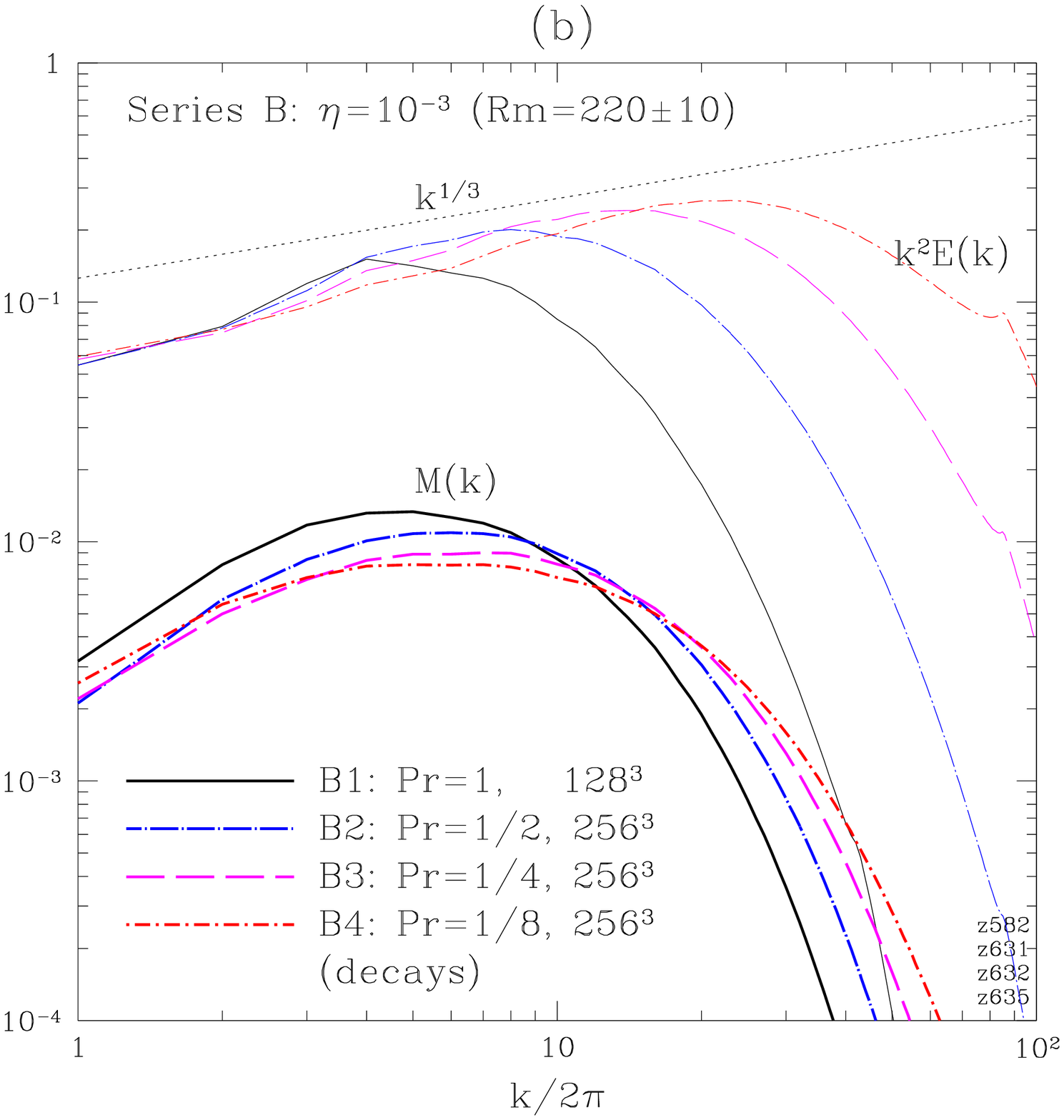,width=8.25cm}}
\caption{\label{fig_Mk} 
Magnetic-energy spectra normalized by $\Bsq/2$ and averaged 
over time: (a) Series~A, (b) Series~B. 
Also given are velocity spectra multiplied by $k^2$ 
and the reference Kolmogorov slope~$k^{1/3}$. 
The time intervals used for averaging are 
for Runs~A1, A2: $5\le t\le40$;  
for Run~B1: $2.5\le t\le17.5$; 
for Runs~A3, B2, B3, B4: $10\le t\le25$.} 
\end{figure*}

In order to address this question, we have carried out a series of 
numerical simulations. 
\Eqsdash{NSEq}{ind_eq} 
were solved in a triply periodic box by the pseudospectral 
method. We used a random nonhelical forcing~$\vf$ applied at the box scale 
and white in time. 
The average injected power 
$\epsilon=\la\vu\cdot\vf\ra$ was kept fixed. 
The code units are based on box size~1 and $\epsilon=1$.
Defining $\Rm=\usq^{1/2}/\eta\kf$ 
with $\kf=2\pi$ the box wave number, we have found that 
the dynamo existed for $\Pr\ge1$ provided 
$\Rm\gtrsim80$ 
(cf.~\figref{fig_Et}b). 
Once the existence of the dynamo at $\Pr=1$ 
for a given $\eta$ was ascertained, $\eta$ was fixed. 
The viscosity~$\nu$ was then decreased. 
The results are summarized in \figref{fig_Et}. 
In runs with~$\eta=2\times10^{-3}$ 
(series A, $\Rm\simeq110$), 
starting from weak field, the dynamo 
persisted at~$\Pr=1/2$, but was shut down at~$\Pr=1/4$. 
In runs with~$\eta=10^{-3}$ 
(series B, $\Rm=220\pm10$), 
there was dynamo at $\Pr=1/2$ and $1/4$, but not at $\Pr=1/8$. 
\Figref{fig_Mk} shows the time-averaged 
normalized magnetic-energy spectra for series A and B, as 
well as velocity spectra multiplied by~$k^2$. The latter 
characterize the turbulent rate of strain and 
peak at the viscous scale. We see that as this scale 
drops below the resistive scale, the dynamo shuts down. 
Note that there is no initial-condition dependence: 
Runs~A3 and~B4, which decay starting from weak field, 
also decay if initialized in the saturated state of their 
$\Pr=1$ counterparts (\figref{fig_Et}a). 

Our main result is, thus, that $\Prc$ exists even as 
$\Rm$ is kept approximately fixed at a value for which 
small-scale dynamo is possible at larger~$\Pr$ (\figref{fig_Et}b). 
In other words, 
{\em the critical magnetic Reynolds number for growth 
{\rm $\Rmc$} increases with} $\Re$. 
Because of resolution constraints, 
we cannot afford a parameter scan to produce the dependence $\Rmc(\Re)$.
What {\em a priori} statements about this dependence 
can be made on physical grounds? 

\begin{figure}[t]
\centerline{\psfig{file=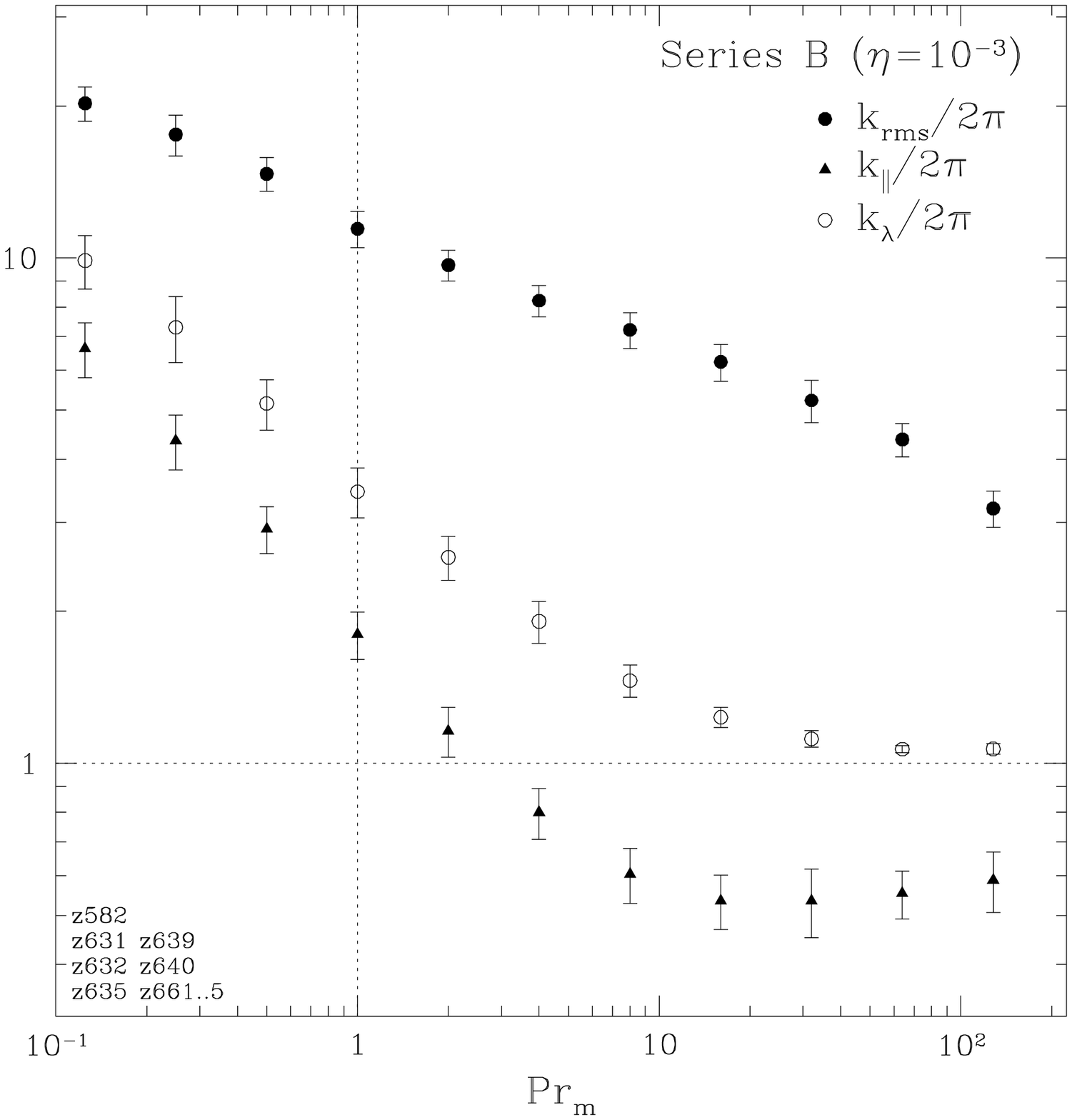,width=8.25cm}}
\caption{\label{fig_k} Characteristic scales for 
Series~B (cf.~\figref{fig_Et}b): 
$\krms=\bl(\la|\nabla\vB|^2\ra/\Bsq\br)^{1/2}$ is 
(roughly) the inverse direction-reversal scale, 
$\kpar=\bl(\la|\vB\cdot\nabla\vB|^2\ra/\Bfr\br)^{1/2}$ 
is the inverse fold length, 
and $\kl=\bl(\la|\nabla\vu|^2\ra/\usq\br)^{1/2}$ is the inverse 
Taylor microscale (times~$\sqrt{5}$) 
(see Refs.~\onlinecite{SCMM_folding,SCTMM_stokes} for 
discussion of these quantities).} 
\end{figure}

Consider first the asymptotic case $\Re\gg\Rm\gg1$. 
The resistive scale then lies in 
the inertial range, $\lf\gg\lres\gg\ld$. 
As, in Kolmogorov turbulence, $u_\ell/\ell\sim\ell^{-2/3}$, 
most of the stretching is done by the eddies 
at the resistive scale $\lres\sim\Rm^{-3/4}\lf$, 
where stretching is of the same order 
as diffusion \cite{Moffatt_lowPr,Vainshtein_lowPr}. 
Since the inertial range is self-similar, the existence of 
the dynamo should not depend on the exact location of~$\lres$, 
and it is the local (in $k$~space) properties 
of the turbulent velocity that determine its propensity to amplify 
magnetic energy. Therefore, if the dynamo fails, it does so 
at all $\Pr$ below some critical value $\Prc$ of order unity. 
The effective transition from the ``large-$\Pr$'' to the 
``small-$\Pr$'' regime occurs at $\Pr=\Prc$. 
In this case, $\Rmc/\Re\to\Prc=\const<1$ as $\Re\to\infty$. 
Thus, if our results are asymptotic, 
then the turbulent small-scale dynamo is always, in essence, 
a large-$\Pr$ one, and the folded direction-reversing 
fields are the only type of magnetic fluctuations that 
can be self-consistently generated and sustained by nonhelical 
turbulence. As the separation between parallel and transverse 
scales of the field diminishes at $\Pr<1$ (\figref{fig_k}), 
no steady fluctuation level can be maintained. 

The second possibility is that $\Rmc$ asymptotes to
some constant value for $\Re$ above those we are able to resolve: 
$\Rmc\to\const\gtrsim220$ and $\Prc\to0$ as $\Re\to\infty$. 
Our results do not rule out this outcome, 
whereby asymptotically in $\Rm$ and $\Re$, the dynamo 
persists at low $\Pr$, but very large $\Rm$ 
is needed to achieve it in practice (numerically 
or experimentally). In stellar convective zones, 
$\Rm$ is, indeed, very large ($10^{6}...10^{9}$ for the Sun). 
On the other hand, in planets and in laboratory dynamos, 
$\Rm\sim10^2$, which is comparable to $\Rm$ in our simulations. 

Note that the arguments above 
assume scale invariance of the inertial range, 
i.e., neglect the effects of intermittency. 
An intermittent velocity field will exhibit large 
coherent fluctuations of the rate of strain, which might 
be locally effective in stretching the magnetic field 
in a way similar to the large-$\Pr$ dynamo \footnote{This point was 
made to us by M.~Vergassola.}. 
Whether these fluctuations can provide enough stretching 
on the average to make a workable dynamo cannot be 
settled qualitatively. Note that an intermittent growth 
by rare strong bursts is evident in $\Pr<1$ runs where the 
dynamo is suppressed but not shut down (most vividly 
in Run~A2, see \figref{fig_Et}a). 

No theory of the dynamo shutdown at low~$\Pr$ 
exists at present. Invoking turbulent diffusion (mixing) 
of magnetic fields by the subresistive-scale motions as the 
suppression mechanism makes heuristic sense, but does not provide 
an unambiguous verdict on the existence of the dynamo.
Indeed, when $\lf\gg\lres\gg\ld$, the dominant contributions to 
both stretching and mixing are from the resistive scale $\lres$, 
and the outcome of their competition is 
impossible to predict on a heuristic basis. 
Many aspects of small-scale dynamo have received successful 
theoretical treatment in the framework of the Kazantsev
model \cite{Kazantsev}, 
which assumes a Gaussian white-in-time velocity field, 
$\la u^i(t,\vx)u^j(t',\vx')\ra = \delta(t-t')\kappa^{ij}(\vx-\vx')$. 
The correlator can be expanded, 
$\kappa^{ij}(\vy)-\kappa^{ij}(0)\sim -y^{\alpha}$,  
when $y\sim\ell_B$, the magnetic-field scale. 
When $\Pr\gg1$, $\ell_B\ll\ld$, so $\alpha=2$, corresponding to 
the spatially smooth viscous-scale eddies. 
On the other hand, when $\Pr\ll1$, $\ell_B\sim\lres\gg\ld$ 
and magnetic field interacts with rough inertial-range velocities, 
which must be modeled by $\alpha<2$.
The Kazantsev velocity is not a dynamo 
if it is too spatially rough, viz., when $\alpha<1$ 
\cite{Kazantsev,Vergassola,Rogachevskii_Kleeorin,Vincenzi}.
If we set aside fundamental objections to the $\delta$-correlated 
model and try to compare it to real turbulence, 
we still face the difficulty of interpreting the $\delta$ function. 
If we write equal-time 
velocity correlators by replacing the $\delta$ function by 
inverse correlation time $1/\tcorr$, then the relation 
between $\alpha$ and the spectral 
exponent of the turbulence depends on how we choose~$\tcorr$. 
The usual choice for Kolmogorov turbulence is 
$\tcorr\sim y^{2/3}$, the eddy-turnover time. 
Then $\alpha=4/3>1$ and there is 
dynamo \cite{Vainshtein_lowPr}\footnote{Note that the EDQNM closure, 
which effectively makes the same assumption 
about the correlation time, gives dynamo at low $\Pr$, 
with $\Rmc$ independent of $\Re$ 
[J.~L\'eorat et al., J.~Fluid Mech.\ {\bf 104}, 419 (1981)].}. 
Note that $\Rmc$ in this case is typically much larger than 
for $\alpha=2$ \cite{Rogachevskii_Kleeorin,Vincenzi,Boldyrev_Cattaneo}. 
Although specific values of $\Rmc$ calculated from the 
Kazantsev model cannot be considered as quantitative predictions 
for real turbulence, they appear to point to the second 
possibility mentioned above (finite but unresolvably large $\Rmc$). 
We emphasize that all these results depend on the heuristic 
choice of~$\tcorr$ (e.g., if $\tcorr\sim\const$, $\alpha=2/3$ 
and there is no dynamo) and on the universality of the 
physically nonobvious condition $\alpha>1$. 
It is fair to observe that our simulations 
are still too viscous to have a well-developed Kolmogorov 
scaling (\figref{fig_Mk}). 
Thus, if the existence of the dynamo depends on the exact 
inertial-range scaling of the velocity field 
and/or only manifests itself at very large $\Rm$, neither 
the Kazantsev theory nor simulations at current resolutions 
can lay claim to a definitive answer. 
Obviously, the Kazantsev theory also cannot capture 
any role the intermittency of the velocity field might play 
and, more generally speaking, it is doubtful that a 
$\delta$-correlated Gaussian flow is a suitable model 
of the inertial-range turbulence. 

While, as far as we know, ours 
is the first systematic study of the small-scale dynamo 
suppression in homogeneous isotropic MHD turbulence with low~$\Pr$, 
indications of this effect have been reported in the literature 
in two previous instances. 
Dynamo suppression at low~$\Pr$ was seen by 
Christensen et al.\ \cite{Christensen_Olson_Glatzmaier} 
in their simulations of convection in a rotating spherical 
shell and by Cattaneo \cite{Cattaneo_lowPr} in simulations of 
Boussinesq convection. 
These cases of failed low-$\Pr$ dynamos 
in inhomogeneous convection-driven turbulence 
are likely to be related to the same universal mechanism 
that made the dynamo inefficient in our simulations. 
Our key conclusion is that the dynamo 
suppression is a generic effect unrelated to the particular 
type of driving or other large-scale features of the system. 

The mean-field dynano, which, in contrast, does depend on 
large-scale features such as helicity and rotation 
\cite{Moffatt_book}, may then be the only type of self-sustained field 
generation for low-$\Pr$ systems.
If a mean field is present, 
it gives rise to a source term~$\Bmean\cdot\nabla\vu$ 
in the induction equation~\exref{ind_eq} 
and thus induces small-scale magnetic fluctuations. 
They have a $k^{-11/3}$ spectrum 
at $k\gg\kres$ \cite{Golitsyn,Moffatt_lowPr,Vainshtein_lowPr}, 
which has been seen in the laboratory \cite{Bourgoin_etal} 
and in large-eddy simulations \cite{Ponty_Pinton_Politano}. 

\begin{acknowledgments}
We thank S.~Boldyrev and F.~Cattaneo for discussions of our results 
and of the Kazantsev model, 
and M.~Vergassola for very useful comments.  
Our work was supported by 
PPARC (PPA/G/S/2002/00075), 
UKAEA (QS06992), 
NSF (AST 00-98670). 
Simulations were run at UKAFF (Leicester) and NCSA (Illinois). 
\end{acknowledgments}

\bibliography{scmm_PRL2}

\end{document}